\documentclass[conference]{IEEEtran}
\IEEEoverridecommandlockouts
\usepackage{cite}
\usepackage{amsmath,amssymb,amsfonts}
\usepackage{algorithmic}
\usepackage{graphicx}
\usepackage{textcomp}
\usepackage{xcolor}
\usepackage{booktabs}
\usepackage{array}
\usepackage{url}
\usepackage{tikz}
\usetikzlibrary{positioning}

\def\BibTeX{{\rm B\kern-.05em{\sc i\kern-.025em b}\kern-.08em
    T\kern-.1667em\lower.7ex\hbox{E}\kern-.125emX}}

\newif\ifextended
\newif\ifshortver
\shortverfalse
\extendedtrue
\newcommand{\extended}[1]{\ifextended \ifshortver \textcolor{purple}{#1} \else \textcolor{black}{#1} \fi \fi}
\newcommand{\shortver}[1]{\ifshortver \ifextended \textcolor{blue}{#1} \else \textcolor{black}{#1} \fi \fi}
\begin{document}

\title{ANI-Gamut: Benchmarking Agent Reliability\\
across the Gamut of\\
Agent--Network Interface Abstractions}

% TODO: confirm author list, order, and affiliations (CNIT? others?).
% Single centered author block (no \and) so the extended-version banner centres full-width.
% Wim Henderickx (Nokia): co-author confirmed; confirm exact unit/country for camera-ready.
\author{\IEEEauthorblockN{Lorenzo Bracciale\IEEEauthorrefmark{1}, Pierpaolo Loreti\IEEEauthorrefmark{1}, Andrea Mayer\IEEEauthorrefmark{1}, Stefano Salsano\IEEEauthorrefmark{1}, Wim Henderickx\IEEEauthorrefmark{2}}
\IEEEauthorblockA{\IEEEauthorrefmark{1}University of Rome Tor Vergata, Italy, \{name.surname\}@uniroma2.it\\
\IEEEauthorrefmark{2}Nokia, Belgium, wim.henderickx@nokia.com}
\ifextended\\[6pt]\large\textbf{Extended version of a paper accepted at CNSM 2026}\fi}

\maketitle

% --------------------------------------------------------------------
\begin{abstract}
% [DRAFT abstract -- ~180 words. The result sentence states the thesis; confirm the exact effect with the measurement campaign.]
Large Language Model (LLM) agents are increasingly trusted to operate live networks: they read state, change configuration, and verify the result.
A first-order question is left implicit: \emph{at which level of abstraction} should the agent operate?
We make the \textbf{interface-abstraction level} an explicit, controlled experimental variable, organizing agent--network interfaces into a spectrum from raw CLI (A0) through bounded wrappers (A1) and standardized model-driven configuration (A2) to typed transactional service intent (A3-T), reconciled source-of-truth automation (A3-R), and their combination (A4).
We present \textbf{ANI-Gamut}, a reproducible, open-source playground that exposes the \emph{same} task at several levels on a single, densely populated \emph{brownfield} substrate, where many coexisting services share resources and collateral damage actually arises.
We instantiate and measure four points of the spectrum and describe the others only at the conceptual level, and record three dependent variables as the substrate is stressed by injected faults: task reliability, collateral damage against pre-existing tenants, and operational cost.
In a pilot with small run counts ($n{=}20$, $n{=}6$ and $n{=}4$ per level), the interface level moves reliability and cost sharply: on a live change-set task a raw-shell agent fails on all six seeds while a typed transactional interface succeeds on all six (paired McNemar $p{=}0.031$, on six discordant pairs), at roughly an order of magnitude less cost, as the engineering effort migrates from the agent to a reusable transaction layer.
Collateral damage, by contrast, is absent at every level, whether benign or under faults: in a tenant-isolated substrate the agents fail safe, and the blast radius is held by the substrate's isolation, which moves the safety question from the agent to the substrate.
ANI-Gamut realizes and extends the benchmarking-playground vision of~\cite{salsano2026playground}, turning the interface-abstraction level from an implicit design choice into a measured variable of agent reliability and cost.
\end{abstract}


% ====================================================================
\section{Introduction}
\label{sec:intro}
% [~1.25 pg] Intro. Source: research plan sec. 1 (research question, 3 contributions, positioning, scope split).
\IEEEPARstart{N}{etwork} operations are moving from human-typed commands to LLM agents that act on live infrastructure, observing the running network, changing it, and confirming the result.
Attention has gone almost entirely to the agent, its model, prompting, and planning loop, while a prior question stays implicit: \emph{at which level of abstraction} should the agent operate?
The same change can be presented to the agent as a raw device shell, as a typed declarative service request, or as an edit to a reconciled source of truth, and these are very different contracts between the agent and the network.

Software engineering has already learned that this choice is first-order.
How reliably a coding agent works is governed by the agent--computer interface, not the model alone.
A constrained, well-shaped interface lets a given model succeed where a raw shell makes it fail~\cite{yang2024sweagent,codestruct2026}.
Networking has begun to borrow the vocabulary, and recent agent--network interface (ANI) work already uses the term, but at a single, fixed abstraction~\cite{cui2025netconfbench}; no study varies the level and measures what it buys.
The question is not academic.
Operators and vendors are climbing an abstraction ladder today, from device CLI to model-driven configuration (NETCONF/YANG~\cite{rfc6241,rfc8342}), to service-intent orchestration (Cisco NSO, Nokia EDA, Juniper Apstra~\cite{ciscoNSO,nokiaEDA,juniperApstra}), to reconciled source-of-truth automation (NetBox/Nautobot~\cite{nautobot}).
Which rung is safe to hand to an autonomous agent is being decided in practice, and it is the decision this paper turns into a measurement.

What makes the question sharp is the failure mode that matters in operations.
A configuration agent that completes its task but silently breaks an unrelated, already-deployed service has, in practice, caused an outage.
Deployability therefore has two legs: the agent must reach its goal \emph{and} leave every pre-existing service intact, the latter at carrier-grade reliability.
Collateral damage is a zero-tolerance property, so we score not only whether the agent finishes but whether a run leaves the network damage-free, and how both hold up as the surrounding state grows denser and faults intrude.

We make the \textbf{interface-abstraction level} an explicit, controlled experimental variable.
We organize agent--network interfaces into a spectrum, A0 to A4 (Fig.~\ref{fig:spectrum}), from the raw shell (A0) through bounded per-device wrappers (A1) and standardized model-driven configuration (A2) to a two-dimensional top: typed transactional service intent (A3-T), reconciled source-of-truth automation (A3-R), and their combination (A4).
Holding the task, the network, the model and the observation interface fixed, we change only the level the agent acts through, and read off reliability, safety, recovery and cost.

Measuring this well needs a substrate where collateral damage can actually occur.
That calls for a \emph{brownfield}: a network already densely populated with heterogeneous services that share resources, so that a careless change can shadow a prefix, reuse an identifier, or trip a global knob.
Toy playgrounds with a handful of services hide the very interactions we want to stress, and a conclusion drawn at toy scale need not hold at production scale.
The machine-learning fields that matured did so on shared benchmarks whose difficulty matched the real problem, such as ImageNet for image recognition~\cite{imagenet} and the WMT shared tasks for machine translation~\cite{wmt2014}; agentic network configuration has no equivalent yet, and constructing one is part of the work.

This paper makes three coupled contributions.
\begin{itemize}
\item \emph{The interface level as a controlled variable.} We define the A0--A4 abstraction spectrum and use it as the independent variable in a controlled study of agent reliability, tracing where each safety property enters. The principle is established in software engineering but not yet in networking under controlled conditions.
\item \emph{A reproducible brownfield playground.} We build, and release as open source, ANI-Gamut: a single multi-service substrate, a layered scenario description, and a seeded initial-state generator that populates a realistic brownfield, on which the same task is exposed at several levels. It instantiates the benchmarking playground envisioned in~\cite{salsano2026playground}.
\item \emph{A methodology for the production-readiness gap.} Damage-free operation is a near-zero-probability, zero-tolerance target, and estimating it is not straightforward. The method combines a binary per-run safety outcome, proportion confidence intervals, run-budget sizing by the rule of three, paired comparisons over shared fault seeds, and stress sweeps over state density and fault rate. This lets an agent's collateral-damage probability, and its distance from the production regime, be estimated and compared across interface levels.
\end{itemize}

Existing agentic-network benchmarks differ from ours on two axes: they fix the agent--network interface as one harness, and they populate little coexisting configuration state, mostly to score troubleshooting and root-cause analysis~\cite{wang2025nika,netopsbench2026,netarena,aiopslab2025,cloudopsbench2026,meshagent2026}.
We instead treat the interface level as the experimental treatment, and push toward realistic brownfield state heterogeneity and density for \emph{safe configuration}, not repair.
ANI-Gamut ships at CNSM as a first reproducible release (the composition manifest, the generator, the harness); a companion extended version, on the project page, details the methodology and the system architecture.\footnote{Playground, code, and the extended version: \url{https://netgroup.github.io/ani-gamut/}.}

In a pilot on a multi-tenant SRv6 brownfield the effect is sharp: at a matched budget, on a live change-set task a raw-shell agent fails on all six seeds where a typed transactional interface succeeds on all six, at an order of magnitude less cost.
Collateral damage, in contrast, stays absent at every level: the tenant isolation of the substrate bounds the blast radius, a result we return to.
The study measures four points of the spectrum (A0, A1, A1+ and A3-T); two points stay conceptual (A2, A3-R) and one is future work (A4), so what the data supports is a contrast between the ends of the implemented range, not a curve over the whole spectrum.

The rest of the paper is organized as follows.
Section~\ref{sec:related} reviews agent interfaces and network benchmarks; Section~\ref{sec:spectrum} defines the abstraction spectrum; Sections~\ref{sec:design} and~\ref{sec:playground} describe the framework and the brownfield playground; Section~\ref{sec:method} states the methodology; Section~\ref{sec:results} reports results; and Sections~\ref{sec:discussion} and~\ref{sec:conclusion} discuss implications and conclude.

% ====================================================================
\section{Background and Related Work}
\label{sec:related}

\subsection{Agent--computer and agent--network interfaces}
The software-engineering community has shown that the interface exposed to an LLM agent, and not only the model behind it, governs how reliably the agent acts.
SWE-agent introduces the \emph{Agent--Computer Interface} (ACI) and shows through ablations that replacing a raw shell with a small set of agent-tailored actions changes task-resolution rates by a wide margin~\cite{yang2024sweagent}.
CODESTRUCT reports the same effect for code editing: moving from free-text patches to structured operations over syntax-tree entities sharply reduces invalid edits~\cite{codestruct2026}.
There is a safety counterpart to this reliability story: granting an agent executable tools rather than text alone measurably raises the rate of unsafe actions under identical prompts and policies~\cite{toolaffordance2026}.
In networking, the term \emph{Agent--Network Interface} (ANI) and the use of the Model Context Protocol (MCP) to expose network operations as tools were introduced by the IETF NetConfBench draft~\cite{cui2025netconfbench}.
We port the ACI principle to networking and add what the software-engineering setting does not address: operational safety and recovery on a live network.

\subsection{LLM agents and benchmarks for network management}
Several benchmarks evaluate LLM agents on network tasks.
NetConfEval asks whether models can translate requirements into formal specifications, function calls, and low-level configuration, and already observes that higher-abstraction targets are easier than raw configuration~\cite{wang2024netconfeval}; it studies these as static translation tasks, not as an agent acting in a closed loop.
NIKA~\cite{wang2025nika} and NetOpsBench~\cite{netopsbench2026} are interactive arenas for troubleshooting and root-cause analysis on emulated networks, scoring detection and localization through bounded observability tools.
The Confucius framework runs multi-agent LLMs in production and routes actions through validated intermediates rather than raw configuration~\cite{confucius2025}.
Related efforts benchmark autonomous remediation in cloud and microservices settings~\cite{microremed2025} and survey the design and safety of agentic NetOps and AIOps~\cite{netopsaiops2026}.
Mani \emph{et al.} make a complementary point outside benchmarking: having an LLM emit code against high-level APIs, instead of touching raw data, improves correctness and explainability~\cite{mani2023}.
Work at CNSM and TNSM has taken the intent route: policy generation from natural language for intent-based application management~\cite{dzeparoska2023}, flow-rule generation for SDN with retry-based deployment validation~\cite{elhachimi2025}, and LLNet, which instructs softwarized devices from intents through a small language model~\cite{angi2025llnet}.
Our novelty is not intent-based management, which that line already pursues, but the controlled comparison of the abstraction levels at which an agent is allowed to act.
In all of this work the interface through which the agent acts is fixed as a single harness; none treats it as an experimental variable.

\subsection{Industrial network automation and autonomy levels}
\label{subsec:industry}
Our premise is that operator networks are already climbing an abstraction ladder, so an AI agent will increasingly have to act \emph{through} these layers rather than around them.
Standardized model-driven configuration is mature: NETCONF/YANG with candidate datastores and confirmed commit~\cite{rfc6241,rfc8342}, RESTCONF~\cite{rfc8040}, and the vendor-neutral OpenConfig/gNMI stack~\cite{openconfig}.
Above the device, intent-based networking~\cite{rfc9315} and autonomic networking~\cite{rfc8993} define declarative, outcome-driven control, while source-of-truth--driven automation~\cite{nautobot} and commercial controllers such as Cisco NSO~\cite{ciscoNSO}, Nokia SR~Linux/EDA~\cite{nokiaEDA}, and Juniper Apstra~\cite{juniperApstra} reconcile a versioned desired state against live device state.
Liu \emph{et al.} organize policy languages by their level of abstraction along the intent-refinement pipeline~\cite{liu2025policyabstraction}.

Operator and vendor bodies have also \emph{classified} how far this automation has progressed.
The TM~Forum Autonomous Networks programme defines six levels, from Level~0 (manual) to Level~5 (fully autonomous), with an evaluation methodology (IG1252) that scores five cognitive dimensions, Intent/Experience, Awareness, Analysis, Decision, and Execution, as owned by personnel or by the system~\cite{tmf_ig1252}.
3GPP SA5 standardizes an equivalent grading in TS~28.100~\cite{3gpp_ts28100}, and ETSI ZSM specifies the closed-loop, intent-driven architecture these levels assume~\cite{etsi_zsm}.
These frameworks classify \emph{autonomy}: how much of the decision loop the system, rather than a human, owns.
Our spectrum (Section~\ref{sec:spectrum}) measures a different and complementary quantity, the abstraction of the interface the agent acts \emph{through}.
The two axes are orthogonal: an agent at any autonomy level still acts on the network through some interface-abstraction level.
To avoid clashing with these autonomy levels, and with OSI layer numbering where L2 and L3 already carry fixed meanings, we denote our interface-abstraction spectrum A0--A4 (Section~\ref{sec:spectrum}) and leave ``L'' to its established uses.

\subsection{The gap}
Two gaps cut across this work.
First, the agent--network interface is always a single fixed harness, whether a typed MCP toolset, an agent--cloud interface, or an emulated CLI~\cite{wang2025nika,netopsbench2026,aiopslab2025}: the interface abstraction level itself is never treated as a controlled variable, held constant across task, network, and model and correlated with reliability, safety, and recovery.
NetConfEval~\cite{wang2024netconfeval} does compare abstraction levels, but as independent static translation tasks (requirements to specification to low-level configuration), not as one task an agent carries out through interfaces held constant and varied across runs on a live network; the IETF NetConfBench draft~\cite{cui2025netconfbench} coins the term ``agent--network interface'' yet fixes it at a single device-CLI abstraction.
The closest a benchmark comes to manipulating the interface is OperAID~\cite{operaid}, whose tool-access ablation moves a 5G-core remediation agent from $11\%$ to $61\%$ success; but it toggles read-only \emph{observation} tools at a fixed raw action interface, confirming that the interface matters while leaving the abstraction level of the agent's \emph{actions}, the variable we study, untouched.

Second, what existing benchmarks under-populate is not device count but the global configuration state that makes configuration hard.
A production backbone or datacenter carries, at the same time, tens to about $10^2$ heterogeneous service classes and on the order of $10^5$ to $10^6$ configuration items in total; it is the size and heterogeneity of this coexisting state, not the number of nodes, that produces the resource sharing and contention where collateral damage occurs.
Existing arenas populate far less of it, and even NETPRESS's~\cite{netarena} thousands of nodes form a single-purpose static capacity-planning graph rather than a dense, heterogeneous live configuration.
ANI-Gamut does not reach production scale either, but it approximates it from below far more closely than toy substrates: on a modest fabric (a few tens of nodes) we populate on the order of ten coexisting service classes (SRv6 east-west and north-south tunnels, firewall policy, underlay routing, floating-IP, addressing, VRFs), each with $10^3$ to $10^4$ rules or configuration items.
\extended{Appendix~\ref{app:realscale} gathers the public evidence behind these orders of magnitude, and Appendix~\ref{app:arenas} compares existing benchmarks on this state-complexity metric.}

Our own prior playground~\cite{salsano2026playground} supplies the starting point: the SRv6 substrate and the typed transactional interface that becomes A3-T here.
What is new in ANI-Gamut is the interface level as a controlled variable, the layered brownfield generator with its co-emitted oracle, and the zero-tolerance damage metric with the statistics that go with it.

By treating the interface abstraction level as the independent variable on this dense, multi-service brownfield substrate, ANI-Gamut closes both gaps.

% ====================================================================
\section{The Interface-Abstraction Spectrum}
\label{sec:spectrum}
\begin{figure}[t]
  \centering
  \includegraphics[width=\columnwidth]{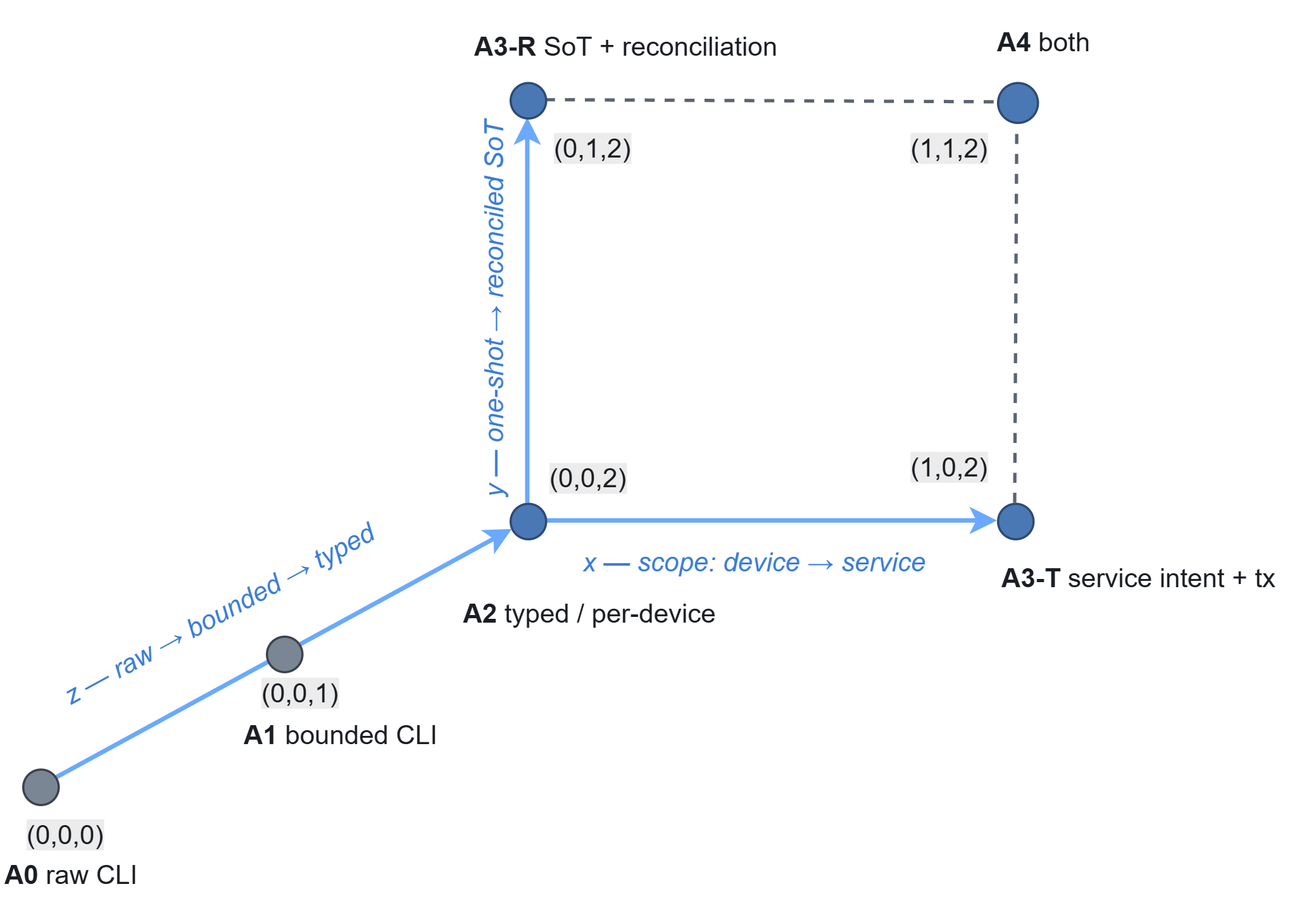}
  \caption{The interface-abstraction spectrum. The ordered base A0--A1--A2 runs
  along $z$ (representation maturity); the high-abstraction interfaces form a
  2D top (Axis~A: scope; Axis~B: state/control), with A2 as origin, A3-T and
  A3-R as the two incomparable corners, and A4 their combination.}
  \label{fig:spectrum}
\end{figure}

We organize agent--network interfaces along a spectrum, shown in Fig.~\ref{fig:spectrum}, from the rawest device access to fully reconciled source-of-truth automation.
The spectrum is the conceptual contribution of this paper: it names \emph{where} an agent acts, independently of which model or agent architecture is used.
Table~\ref{tab:spectrum} summarizes the six levels and the aligned subset ANI-Gamut evaluates.

\begin{table}[t]
\caption{The interface-abstraction spectrum and the subset ANI-Gamut evaluates (\textsc{impl.}\ implemented, \textsc{conc.}\ conceptual, \textsc{fut.}\ future work).}
\label{tab:spectrum}
\centering
\footnotesize
\setlength{\tabcolsep}{4pt}
\begin{tabular}{@{}llp{3.2cm}c@{}}
\toprule
\textbf{Level} & \textbf{Interface} & \textbf{Key safety / transaction property} & \textbf{Eval.} \\
\midrule
A0    & raw shell      & none; unbounded action space                     & impl.  \\
A1    & bounded CLI    & generic guardrail $+$ manual recovery            & impl.  \\
A2    & NETCONF/YANG   & device candidate commit $+$ validation           & conc.  \\
A3-T  & service intent & workflow transaction $+$ validation $+$ rollback & impl.  \\
A3-R  & reconciled SoT & drift detection $+$ reconciliation               & conc.  \\
A4    & intent $+$ SoT & both high-level axes (Pareto-dominant)           & fut.   \\
\bottomrule
\end{tabular}
\end{table}

\subsection{The ordered base: A0 to A2}
At the bottom of the spectrum the agent operates on individual devices.
\textbf{A0} is the raw device shell, an unbounded action space with no transactional semantics.
\textbf{A1} wraps per-device configuration behind a bounded interface (read the running configuration, apply a change, run a validation command), still expressed as free-text commands.
\textbf{A2} is standardized, model-driven configuration over NETCONF/YANG: typed and vendor-neutral, with candidate datastores and confirmed commit that provide device-level transactional safety~\cite{rfc6241,rfc8342}.
These three levels are ordered by how far the interface moves the agent from raw mutation and by how much safety it interposes.

\subsection{The two-dimensional top: A3-T, A3-R, and A4}
Above A2 the spectrum is not a single ladder but a plane spanned by two orthogonal axes.
The first axis is \emph{scope}, from per-device configuration to service-level, cross-device intent.
The second is the \emph{state and control model}, from one-shot, episodic actions to a persistent, versioned source of truth continuously reconciled against live state.
\textbf{A3-T} (transactional service intent) advances the scope axis: the agent issues a typed declarative service request that a workflow transaction manager applies with active validation and rollback; this is the interface realized by the SRv6 playground~\cite{salsano2026playground}.
\textbf{A3-R} (reconciled source of truth) advances the control axis: the agent edits an authoritative, versioned desired state, and a reconciliation loop detects drift and remediates~\cite{nautobot}.
A3-T and A3-R are not ordered with respect to each other; each provides a different kind of safety.
\textbf{A4} occupies the remaining corner, combining service intent with reconciliation, and is where commercial systems such as Cisco NSO, Nokia EDA, and Juniper Apstra already operate~\cite{ciscoNSO,nokiaEDA,juniperApstra}.
Placing A2 at the origin, A3-T, A3-R, and A4 are the points $(1,0)$, $(0,1)$, and $(1,1)$: A4 Pareto-dominates both A3 variants, which is why it carries the higher number even though A3-T and A3-R are mutually incomparable.

\subsection{Abstraction exposed, not implementation}
A level denotes the abstraction \emph{exposed to the agent}, not the stack used to enforce it, and a level need not be built on the one below.
A3-T does not require A2: the SRv6 playground enforces its typed \texttt{encap}/\texttt{decap} intent through ordinary kernel routing commands, with no NETCONF involved, and a configuration-domain A3-T can likewise be realized by serializing typed intent into device commands.
This decoupling lets us study the interface the agent sees while leaving the realization substrate free, and it is what allows the same task to be exposed at several levels (Section~\ref{sec:design}) on one emulated testbed (Section~\ref{sec:playground}).

% ====================================================================
\section{ANI-Gamut: Levels as a Controlled Variable}
\label{sec:design}
% [~1 pg] FRAMEWORK + PER-LEVEL ADAPTERS. Source: research plan sec. 5/5.1/5.2.

\begin{figure}[t]
  \centering
  \includegraphics[width=\columnwidth]{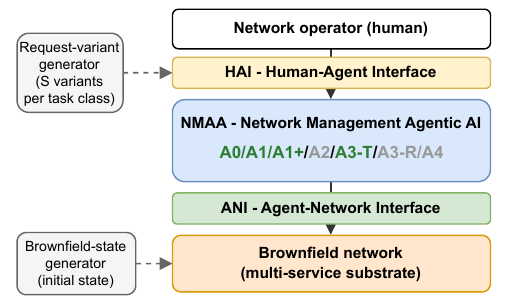}
  \caption{ANI-Gamut at a glance. The NMAA system sits between the human--agent interface (HAI), where the operator states a configuration requirement, and the agent--network interface (ANI), through which it acts. Only the ANI level varies (green: implemented; grey: conceptual); the HAI request, the model and the observation are held constant. Two generators drive each run: the request variants of a task class and the seeded brownfield state.}
  \label{fig:arch}
\end{figure}

ANI-Gamut makes the interface level the one variable that changes between runs.
The system under test is a \emph{network-management agentic-AI} (NMAA) system that sits between two interfaces: it receives a configuration requirement from a human operator over a \emph{human--agent interface} (HAI), a natural-language prompt with zero or more attached files, and acts on the network through the agent--network interface (ANI) (Fig.~\ref{fig:arch}).\extended{ Appendix~\ref{app:arch} expands this with the scenario generators and the oracle.}
The ANI level is an internal choice of the NMAA system, invisible to the operator: the same HAI request is met by A0 with a raw shell and by A3-T with typed intent.
We therefore hold the HAI request, the observation interface and the model fixed and vary only the ANI level, so two NMAA systems compared here are identical except for the abstraction at which they act.
We build the harness by unifying two existing codebases: NetAiBench, a configuration harness developed in our group that already sits at A1, and the SRv6 playground~\cite{salsano2026playground}, which already realizes an A3-T service interface.
The full A0--A4 spectrum is the conceptual contribution; the demonstrator implements the aligned subset A0, A1 and A3-T, leaving A2 and A3-R conceptual for the reasons given in Section~\ref{sec:method}.

\subsection{A unified harness and a common ANI}
The harness derives from NetAiBench's modular architecture.
Every task carries one specification: a natural-language intent, the topology, the expected end state, machine-checkable test cases, and the \emph{baseline invariants} that the pre-existing services and global settings must keep satisfying.
A common ANI sits between the agent and the substrate, with one adapter per level that translates the agent's actions into substrate operations.
The agent is a controlled factor, not the object of study: single-turn, ReAct and planner--executor modes, with and without recovery, run unchanged across levels.
A shared evaluator scores every run identically and observation is held constant, so that what differs between two runs of the same task is only the level the agent acts through.
The substrate is the unified Containerlab testbed of Section~\ref{sec:playground}, on which both source codebases already run.

\subsection{Per-level adapters (same task, different interface)}
At \textbf{A0} the adapter is a raw shell on the node: an unbounded action space with no transactional semantics, the naive path obtained by removing every guardrail.
\textbf{A1} keeps the same free-text representation and adds two separable mechanisms.
The first is \emph{bounding}: a configuration-command allowlist, a block on out-of-scope shell, and protection of the management plane.
The second is \emph{recovery}: an auto-captured per-device snapshot the agent can roll back.
This bounding is a \emph{generic} guardrail, keeping the agent inside the configuration surface and the device reachable, not an anti-damage blocklist; hand-picked prohibitions against specific destructive commands would make the level safe by our own tuning, so we move them into the \textbf{A1+} ablation (Section~\ref{sec:method}).
In the configuration domain A0 and A1 are deliberately close, differing only by the allowlist, since both expose raw CLI; the substantive jump is to typed intent.
\textbf{A3-T} exposes typed declarative \emph{service} intent: the agent issues a cross-device service request that a workflow transaction manager applies with active validation and compensating rollback.
It already exists in the SRv6 playground as schema-checked \texttt{encap}/\texttt{decap} messages with a per-direction transaction and an ICMP-in-SRv6 validation probe~\cite{salsano2026playground}; we generalize it to the configuration domains by promoting NetAiBench's internal typed model and inverse serializers, today used only for rollback, to a public interface.
\textbf{A2} (standardized NETCONF/YANG with candidate datastores and confirmed commit~\cite{rfc6241,rfc8342}) and \textbf{A3-R} (a versioned source of truth reconciled against live state~\cite{nautobot}) are defined but left conceptual: no mature open-source orchestrator gives an honest, measurable A2 agent path on this substrate, and a reconciliation loop is a journal-phase extension.
Attribution does not depend on them; it rests on contrasts we can run, the A0-versus-A1 bounding and recovery decomposition and A3-T-internal ablations, detailed in Section~\ref{sec:method}.

% ====================================================================
\section{A Reproducible Brownfield Playground}
\label{sec:playground}
% [~1.5 pg] BROWNFIELD SUBSTRATE + LAYERED GENERATOR + FAULTS + OPEN RELEASE.
% Source: research plan sec. 5.2/5.3/5.4.

The playground is the substrate where collateral damage can occur, and we release it as a reproducible open-source artifact with the paper: a single multi-service substrate, a layered scenario description, and a seeded initial-state generator.

\subsection{A unified multi-service substrate}
All services coexist on the same devices, as on a real provider edge: SRv6 east-west and north-south tunnels, firewall policy, underlay routing, floating-IP and addressing share each node's namespaces.
That shared occupancy is what makes resource contention, and therefore collateral damage, real.
It is feasible because both source codebases run on Containerlab with Linux and FRR, so their labs merge onto one node image (the SRv6 node, with \texttt{seg6} and FRR, extended with \texttt{nft} and \texttt{iptables}); NetAiBench's SR~Linux and OpenBSD drivers add device heterogeneity.
The SRv6 service class runs on a NEXT-CSID (micro-SID) data plane whose locators follow the F3216 addressing recommendations of the IETF SRv6ops draft~\cite{horn2026srv6addressing} (a single 32-bit block, with 8-bit set and node fields and loopbacks taken from the locator), so the testbed reflects current operator practice rather than an ad-hoc scheme.
Topology and state are two independent scales.
The number of emulated nodes is bounded by host resources, but the pre-existing state (tenants, VRFs, SIDs, routes, rules) is kernel and configuration data, not extra containers, so it grows independently.
Our hypotheses live in the \emph{state}, so we keep a modest fabric, six PEs, three IGWs, two EGWs and twelve hosts, about twenty-five nodes, and populate it densely.
On it we pre-load on the order of ten coexisting service classes with $10^3$ to $10^4$ items each, an under-approximation of the production orders of magnitude of Section~\ref{sec:related}.
State heterogeneity and density are parameterized and swept: the prediction is that the safety advantage of higher abstraction grows with density, hidden on a sparse substrate and revealed on a dense one.

\subsection{A layered, composable scenario generator}
The brownfield is built like a container image.
A topology base, the Containerlab description, carries a stack of service-class layers declared in a composition manifest, a brownfield ``Dockerfile'' written in YAML.
Each service class is a plug-in layer with three phases.
An \texttt{init} phase establishes the class prerequisites, enabling \texttt{seg6}, reserving VRF and SID space, or setting up base firewall chains.
A \texttt{generate} phase writes the class's $T{=}0$ instances as real kernel and device state, together with the baseline invariants for those instances, so the oracle is emitted alongside the state.
A \texttt{verify} phase checks every instance of the class, exhaustively or sampled.
The manifest lists the layers, their dependencies, the per-class parameters (instance counts, prefix overlap, load skew, distribution shapes) and the seed; topology and manifest together determine the scenario.
Phases run globally, \texttt{init} then \texttt{generate} then \texttt{verify}, with layers ordered by dependency, so the underlay routing is in place before the overlays that ride on it.
The oracle is therefore distributed across layers, each class verifying its own instances, and the agent's target is a held-out instance of a task class, declared in the manifest and placed to interact with existing state, on the same PE, an adjacent prefix, or a shared SID pool; the same generator draws the random variants of that class (Section~\ref{sec:method}) that make up a cell's paired runs.
A realistic initial state, rather than a merely dense one, means several things at once.
Every pre-existing service passes its invariants at $T{=}0$, or the scenario is regenerated.
Service sizes and node load follow plausible skewed distributions, shared-resource dependencies and prefix adjacency or overlap occur at a tunable rate so that shadowing and collisions are genuine traps, the per-class mix is controlled, and the target sits in the populated context rather than in isolation.
How much the generator's realism affects the conclusions is itself a question, which the density sweep turns into a journal-phase sensitivity study.

\subsection{Substrate-level fault injection}
Faults are injected in the shared substrate, identically for every level, never inside a per-level adapter.
Injecting ``operation $X$ fails with probability $p$'' at A3-T would bake the level's own resilience into the stimulus and destroy comparability; instead we inject the cause low, below the ANI adapters, and measure how the effect propagates through each interface.
The admissible primitives act at node and link level and strike all levels alike: packet loss, latency and jitter, a downed link, an unresponsive (frozen) node, a daemon restart, a held route-table or datastore lock, a concurrent writer mutating shared state during the task, and transient exhaustion of a shared SID or address pool.
Transport-specific faults are not admissible as primitives, because transports differ by level; they are modeled as node or link faults that hit every transport equally.
Partial application, double-apply on retry, and lockout are measured outcomes, not inputs: they emerge when a primitive strikes at the wrong moment, so the same physical event can leave A0 with half-applied state while an A3-T transaction rolls back or retries idempotently.
Faults are seeded and the same seed is replayed across levels, common random numbers made possible precisely because the fault lives in the substrate; observation is out-of-band, run by the harness rather than the agent's level-specific validation, so success and safety are measured identically everywhere.
The fault rate is a tunable stress knob, set deliberately above realistic levels, and the gap between levels as a function of it is a headline analysis alongside the density sweep.

\subsection{A reproducible open-source release}
Adding a service class is adding a layer with its three phases; the playground composes the layers and derives the oracle automatically.
This realizes and extends the single extensible benchmarking playground envisioned by the accepted SRv6 paper~\cite{salsano2026playground}.
The tooling, the layered scenario description and the initial-state generator with the harness, ships as a first reproducible open-source release with this paper rather than being deferred to the journal; designing it for reuse is part of the contribution.
Large campaigns are made practical by engineering the testbed for scale, through parallel scenario execution and a reduced per-node memory footprint.\extended{ Appendix~\ref{app:scalability} reports the parallelization scheme and the footprint reduction.}

% ====================================================================
\section{Experimental Methodology}
\label{sec:method}

\subsection{Hypotheses}
We test six directional hypotheses, some of which may be refuted; that is what makes this measurement rather than advocacy.
\textbf{H1 (safety):} higher-abstraction interfaces lower the rate of collateral damage relative to the A0/A1 baseline, at constant task and model.
\textbf{H2 (recovery):} higher levels return the network to a safe, correct state more often after a committed error.
\textbf{H3 (expressiveness trade-off):} functional success is not monotonic in abstraction, because very high levels can cap what is expressible on some tasks.
\textbf{H4 (cost):} higher levels reach a correct-and-safe state at lower cost and lower variance.
\textbf{H5 (model $\times$ level):} the benefit of abstraction is larger for weaker models, the interface compensating for model capability.
\textbf{H6 (attribution):} the reliability gain comes from transactional and recovery semantics entering at specific points, not from typed representation alone.

\subsection{Design and controlled factors}
The interface level is the treatment in a factorial design that crosses it with the task class, the model, and the agent mode, swept over two stress parameters: the state density of the brownfield and the substrate fault rate $\lambda$.
A \emph{task class} is a configuration requirement stated on the HAI, a prompt template with zero or more attached files; the task classes span the service classes that populate the substrate, and two classes that request the same reconfiguration in different words isolate the effect of expression from the effect of the situation.
For each cell the generator draws $S$ random \emph{variants} of the task class, each a concrete request (tenant, prefixes, target) placed in a freshly generated brownfield; the prompt text and the file structure are identical across a class's variants, so what varies is the situation, not the wording.
These $S$ variants are the per-cell runs over which the statistics below are estimated, and each variant, bundling the request, the brownfield, and the fault seed, is replayed across the levels, which is what makes the comparison paired.

We compare interface levels under a matched agent budget.
The execution budget is held constant across levels, with the same model, system prompt, agent mode, step and token caps, retry policy, and observation interface, so that only the action interface differs.
The development budget is equalized by a single shared agent with no level-specific tuning beyond exposing each interface.
We additionally report cost (tokens, messages, steps, wall-clock) as a dependent variable (H4): the question is not only whether a higher level is safer at equal budget, but whether it reaches a safe state at lower budget.

The slice implemented here is A0, A1 and A3-T, with A1+ as an ablation, while A2, A3-R and A4 stay conceptual.
This still answers the question: the claim concerns the \emph{trend} from the low-abstraction baseline to the high-abstraction interfaces, not a continuous monotone curve, and attribution is preserved through within-level ablations rather than through the absent A2/A3-R contrasts.

\subsection{Metrics: a zero-tolerance safety outcome}
Agents are usually scored on mean task completion, but moving a success rate from $95\%$ to $97\%$ leaves a configuration agent far from deployable.
A deployable agent must reach a success probability indistinguishable from one \emph{and} damage no pre-existing service at carrier-grade reliability, the ``five nines'' of network practice, a collateral-damage probability on the order of $10^{-5}$.
No current system is near this bar on either leg; the playground measures how far a system is and which knobs move it closer.
This fixes the role of every metric: task completion and the breadth of damage do not rank near-deployable systems, since none exists, but show that any system for which they are non-trivially measurable is by that fact far from production, and chart the directions in which the knobs (level, model, agent mode, stress) move it.

The unit of damage is the \emph{service instance}, one deployed service such as a single SRv6 tunnel, grouped into service classes by type.
An instance is damaged if any of its baseline invariants fails after the agent acts, counted once however many invariants it carries.
Per run we record the per-class damage fraction $B_c$ over the whole gamut of pre-existing services (SRv6 east-west and north-south, firewall, routing, floating-IP) and the global invariants (backbone reachability, forwarding, default-route removal, VRF integrity).
Since collateral damage must be zero, incidence matters more than magnitude: the per-run outcome is binary, the damage vector $\mathbf{B}=(B_c)$ is null or not, and the primary safety statistic over a set of runs is the \textbf{damage-free run rate} $P(\mathbf{B}=\mathbf{0})$.
Magnitude is kept as a secondary, diagnostic measure: the macro-average $\bar{B}$ across classes (equal class weight, insensitive to differing instance counts) and the per-class vector when damage occurs, with pooled micro-averages reported only as an operational footprint.
The functional-success rate $S_{\mathrm{tc}}$, the fraction of test cases the newly configured service passes, is the completion leg, measured but not the safety headline.
Each run also carries an outcome label: clean success, dangerous success (works but damages), clean failure, or destructive failure.
Recovery (H2) uses the same zero-damage lens: the fraction of fault-hit runs that still end damage-free and correct.
Observation is out-of-band, so success and safety are scored identically at every level; the oracle reuses NetAiBench's invariant checks and the SRv6 ICMP-in-SRv6 probe, with the generator co-emitting the per-instance invariants (Section~\ref{sec:playground}).

\subsection{Statistics and attribution}
The primary outcome is binary, so the estimand is a probability: per cell we estimate $\hat{p}(\text{damage})=k/S$ with a Wilson or Clopper--Pearson interval, since the normal approximation fails near zero, which is exactly the A3-T regime.
Sample size follows from the zero claim, not from comparing means: by the rule of three, zero damages in $S$ runs give an upper $95\%$ bound near $3/S$, so asserting $p(\text{damage})<1\%$ for A3-T needs $S$ on the order of a few hundred per cell, and the pilot fixes it.
Common fault seeds are replayed across levels, which makes the comparison paired and lets McNemar's test work on the discordant pairs where the same physical event damaged the low level but not the high one, more direct than comparing marginal rates.
Secondary continuous metrics (cost, time-to-safe-state, magnitude when damage occurs) use confidence intervals and mixed-effects regression with task and model as random effects, including the model$\times$level interaction (H5).
We report success and cost per level with interval estimates.
The density and $\lambda$ sweeps can be found in the extended version.
The hypotheses of Section~\ref{sec:method}-A and the analysis plan are pre-registered and released with the playground.
% TODO(cite): aggiungere \cite{anigamut2026extended} (ID arXiv) sulla frase dei sweep
% e \cite{anigamut2026artifact} (pagina/repo) sulla frase della pre-registrazione,
% appena create le due voci in references.bib.

Attribution (H6) uses contrasts that probe one mechanism at a time, because A3-T and A3-R advance different axes rather than forming one staircase.
A1 bundles \emph{bounding} (allowlist, out-of-scope-shell block, management protection) and \emph{recovery} (snapshot and manual rollback); we vary them as a $2\times2$ factor at constant free-text representation, separating how much \emph{preventing} buys from how much \emph{undoing} buys, which a single A0-versus-A1 contrast conflates.
The A1-to-A3-T step then isolates service-scope typed intent with one-shot workflow transactions.
The way an interface enforces consistency is itself a sub-axis of increasing principledness: an ad-hoc prohibition (the A1+ blocklist), predetermined domain-specific templates, standardized validation and commit (A2), and typed transactional intent (A3-T) approach the same guarantee by different routes, and we compare them empirically instead of assuming the typed route wins.

% ====================================================================
\section{Results}
\label{sec:results}

These results are a pilot: $n{=}20$ per level on the benign task, $n{=}6$ on the change-set, $n{=}4$ under faults; we rely on paired tests and explain the mechanism behind each one.
Two service families are exercised, a benign onboarding task and a change-set on a live tenant, each run with and without injected faults, with one model held fixed across levels and a matched budget (same model, prompt, and step cap; only the action interface changes).
Interface abstraction moves reliability and cost; it does not move the rate of collateral damage.

\textbf{Reliability is governed by abstraction.}
% VOCE 8 (camera-ready 19/9): la tabella tab:onboarding esce solo nella versione
% extended; nella conferenza i numeri restano in prosa. Le due varianti sono
% selezionate dai toggle \shortver / \extended del preambolo.
\shortver{On the benign onboarding task success grows monotonically over the low levels (A0 $0.20$, A1 $0.35$, A1+ $0.55$, $n{=}20$; the Wilson intervals are wide and overlap at this $n$, and the paired McNemar contrast of A0 against A1+ gives $p{=}0.092$, a trend); A1+ above A1 is noise ($p{=}0.289$), since the Tier-2 blocklist can only forbid, not help solve.}
\extended{On the benign onboarding task success grows monotonically over the low levels (Table~\ref{tab:onboarding}); A1+ above A1 is noise, since the Tier-2 blocklist can only forbid, not help solve.}
The separation becomes clean once the high level is in play.
On a live change-set task (two adds, two moves and one remove on a 25-site tenant, $n{=}6$ seeds, Table~\ref{tab:changeset}) A0 solves $0/6$ and A3-T $6/6$, with non-overlapping Wilson intervals and a significant paired test (exact McNemar $p{=}0.031$ on six discordant pairs); A0 reaches about $85\%$ of the mesh on every seed yet never completes it, while A3-T always reaches $100\%$.
The gap survives stress: under injected packet loss and node isolation on the onboarding task ($\lambda{=}0.3$, Table~\ref{tab:faultcal}), A3-T solves $4/4$ where A0 collapses to $0/4$.
The mechanism is that A3-T applies a typed intent through its actuation channel and does not depend on probing the data plane, which the faults have degraded, whereas A0 and A1 must read a broken network to choose each next command, and lose their way.

\textbf{Cost is governed by abstraction.}
Solving through one typed \texttt{provision\_service} call costs far less than composing raw commands step by step: on the change-set, A3-T uses $3$ steps and about $42$k tokens against A0's $40$ steps (capped) and about $375$k tokens, roughly $13\times$ fewer steps and $9\times$ fewer tokens; on onboarding the token cost already rises $+62\%$ from A0 to A1.
Part of the ratio is budget exhaustion rather than inefficiency: A0 is capped at 40 steps and fails at the cap, so its cost is the budget it was given, not the cost of a completed run.
The engineering moves into the layer instead of into each run: the transaction recipe is written and validated once there, so the per-run agent budget falls as the level rises.

\ifextended
\begin{table}[t]
\caption{Capability on the benign onboarding task (no faults; $n{=}20$ per level; one model, matched budget). Wilson intervals are wide and overlap at this $n$. Paired McNemar: A0 against A1+ $p{=}0.092$; A1+ against A1 $p{=}0.289$.}
\label{tab:onboarding}
\centering
\footnotesize
\begin{tabular}{@{}lcccc@{}}
\toprule
\textbf{Level} & \textbf{Success} & \textbf{Rate} & \textbf{Wilson 95\%} & \textbf{Dmg-free} \\
\midrule
A0  & \phantom{0}4/20 & 0.20 & $[0.08,0.42]$ & 1.00 \\
A1  & \phantom{0}7/20 & 0.35 & $[0.18,0.57]$ & 1.00 \\
A1+ & 11/20 & 0.55 & $[0.34,0.74]$ & 1.00 \\
\bottomrule
\end{tabular}
\end{table}
\fi

\begin{table}[t]
\caption{Headline capability on the change-set task (live 25-site tenant; two adds, two moves, one remove; $n{=}6$ seeds, no faults; one model, matched budget). Both levels are damage-free; exact paired McNemar $p{=}0.031$.}
\label{tab:changeset}
\centering
\footnotesize
\begin{tabular}{@{}lccccc@{}}
\toprule
\textbf{Level} & \textbf{Success} & \textbf{Wilson 95\%} & \textbf{Steps} & \textbf{Tokens} & \textbf{Dmg-free} \\
\midrule
A0   & 0/6 & $[0.00,0.39]$ & 40 & ${\sim}375$k & 1.00 \\
A3-T & 6/6 & $[0.61,1.00]$ & \phantom{0}3 & ${\sim}42$k  & 1.00 \\
\bottomrule
\end{tabular}
\end{table}

\begin{table}[t]
\caption{Fault calibration ($\lambda{=}0.3$, onboarding task, $n{=}4$ per level, one model, matched budget). \emph{Success} is the fraction of runs that reach the new service; \emph{damage-free} is the rate $P(\mathbf{B}{=}\mathbf{0})$ of runs with no collateral damage to pre-existing tenants.}
\label{tab:faultcal}
\centering
\footnotesize
\begin{tabular}{@{}lccccc@{}}
\toprule
\textbf{Level} & \textbf{Success} & \textbf{Wilson 95\%} & \textbf{Dmg-free} & \textbf{Steps} & \textbf{Tokens} \\
\midrule
A0    & 0/4 & $[0.00,0.49]$ & 1.00 & 25.0 & 127k \\
A1    & 2/4 & $[0.15,0.85]$ & 1.00 & 25.0 & 228k \\
A1+   & 1/4 & $[0.05,0.70]$ & 1.00 & 25.0 & 239k \\
A3-T  & 4/4 & $[0.51,1.00]$ & 1.00 & \phantom{0}5.5 & \phantom{0}35k \\
\bottomrule
\end{tabular}
\end{table}

\textbf{Collateral damage is \emph{not} governed by abstraction.}
The damage-free rate is $1.00$ at every level, on both tasks, benign and under faults: no run damages a pre-existing tenant.
The null has a clear cause.
The agents fail safe, abandoning the task instead of corrupting their neighbors, and the substrate keeps tenants in separate VRFs, so even overlapping addresses on shared routers cannot leak; the substrate's isolation bounds the blast radius, and the agent's abstraction has no part in it.
What decides collateral safety is where shared state lives, and on this substrate the A1/A1+ blocklist, built to prevent damage, is inert, which is why its ablation reads as noise.
It would bite on services with shared mutable state that tenant isolation does not protect, such as a stateful middlebox with a common ruleset that a careless flush can break; surfacing that is the natural next study.

\textbf{The fault-robustness gap is confounded.}
Part of A3-T's advantage under faults is that its actuation is out-of-band from the data-plane disturbance, so while the actuation channel is the same for every level, the low levels also pay for having to observe a degraded data plane.
Typed representation, transactional semantics, the out-of-band actuation channel and the engineering that went into the adapter vary together here; this pilot does not separate them, and the matched budget equalizes the agent's allowance, not these mechanisms.
The extended version treats the actuation channel as a controlled variable.

% ====================================================================
\section{Discussion}
\label{sec:discussion}

A3-T and A3-R are different bets on safety: A3-T narrows what the agent can express to a typed service request and wraps it in a transaction, while A3-R lets it edit a versioned desired state that a reconciliation loop drives toward.
Our slice tests the first and leaves the second conceptual, so which of them buys more safety is a question the framework is built to answer rather than one we settle here.

The expressiveness trade-off (H3) is the obvious objection to ``higher is safer'': a typed service interface cannot express a configuration its schema does not anticipate, so on tasks such as dynamic-routing tuning a leaner interface that stays close to the device can win on raw success.
We keep such tasks in the corpus precisely so the expressiveness ceiling appears as a result rather than hiding as a gap.

Where does collateral safety come from?
A hand-curated blocklist (A1+) forbids the specific destructive commands an operator has been burned by, while a structural interface makes whole classes of damage unrepresentable; on our substrate neither was exercised, because no level damaged a neighbor.
The data points instead to the substrate: tenant isolation held the blast radius at zero on every run, so on a VRF-isolated brownfield collateral safety is a property of where shared state lives.
The structural-versus-ad-hoc contrast becomes measurable only on services whose shared state the isolation does not cover, which is where we take the study next.

For an operator the question is one of placement, at which rung an agent acts reliably and cheaply on a given service given the model in hand and the stress on the surrounding state, and the framework turns that into the success rate and the budget at each rung rather than a matter of judgment.

The operator--agent interface (HAI) has its own abstraction spectrum, from a terse ticket to a fully structured intent; we hold it fixed so that the agent--network interface is the sole treatment, and leave that symmetric axis to future work.

Several threats temper the conclusions.
The substrate is emulated, so absolute rates will differ on production hardware; we therefore read level-to-level differences, which the paired design protects, rather than absolute values.
The generator argues for realism but does not prove it, and how much its choices move the conclusions is left to the density sweep of the extended version.
A2 and A3-R are conceptual here, so the high end of the spectrum rests on A3-T alone; closing that is the first item of future work.

A subtler concern is the engineering that moves into the shared infrastructure: the transaction manager, active validation, and rollback are built once for all levels.
This is the agent--computer-interface premise at work, and we attribute the resulting reliability to specific mechanisms through the ablations.
The comparison therefore holds at a fixed, modest agent budget; the asymptotic regime, in which unbounded per-level engineering is poured into a low-level agent, is out of scope and operationally irrelevant, since operators do not invest person-years of engineering per agent.

% ====================================================================
\section{Conclusion and Future Work}
\label{sec:conclusion}

We treated the abstraction level of the agent--network interface as an experimental variable and built ANI-Gamut to measure what each level buys.
In this pilot, and between the two implemented ends of the spectrum, the interface level is a measurable determinant of how reliably and how cheaply an agent operates: on a dense brownfield, raising it carries a raw-shell agent from failing to succeeding at a fraction of the cost.
Collateral safety, by contrast, did not move with the interface; on a tenant-isolated substrate it was a property of the isolation model, which relocates the safety question from the agent to where shared state lives.
We still score collateral damage as a binary, zero-tolerance, damage-free run rate, and we supply the statistics, proportion intervals, rule-of-three budgeting, and paired tests, that let a near-zero damage probability be bounded and a small-sample capability gap be tested.
The playground, the layered scenario description, the initial-state generator and the harness, is released as a first reproducible open-source benchmark with this paper.

Future work fills in the spectrum and grows the benchmark.
An implemented A2 (NETCONF/YANG on SR~Linux) and a real A3-R (a reconciliation loop over a source of truth) would complete the high end and let A3-T and A3-R be compared directly; the A4 corner, where commercial systems already operate, follows.
On the measurement side, the extended version quantifies how far undersized playgrounds distort conclusions and opens the release into a shared brownfield challenge.

% ====================================================================
% EXTENDED-ONLY APPENDICES (arXiv version): D1 real scale, D2 arena survey.
% Gated by \ifextended so they never appear in the 9-page conference build.
% ====================================================================

% --------------------------------------------------------------------
% TODO: add bib entries to references.bib (some already present:
%   salsano2026playground, wang2024netconfeval, wang2025nika,
%   cui2025netconfbench, yao2022react). To ADD:
%   yang2024sweagent (SWE-agent, NeurIPS'24), codestruct2026 (ACL'26),
%   netopsbench2026, meshagent2026 (SIGMETRICS'26), microremed2025,
%   toolaffordance2026 (arXiv:2603.20320), agenticpatterns2026 (Electronics),
%   netops_aiops_safety2026 (arXiv:2605.12729), liu2025policyabstraction,
%   rfc9315, terminalbench2026.
\ifextended
\appendices

\section{Detailed System Architecture}
\label{app:arch}
Figure~\ref{fig:arch-detail} expands Fig.~\ref{fig:arch} with the two scenario generators, the per-instance oracle, and the multi-service substrate.

\begin{figure*}[t]
  \centering
  \includegraphics[width=0.72\textwidth]{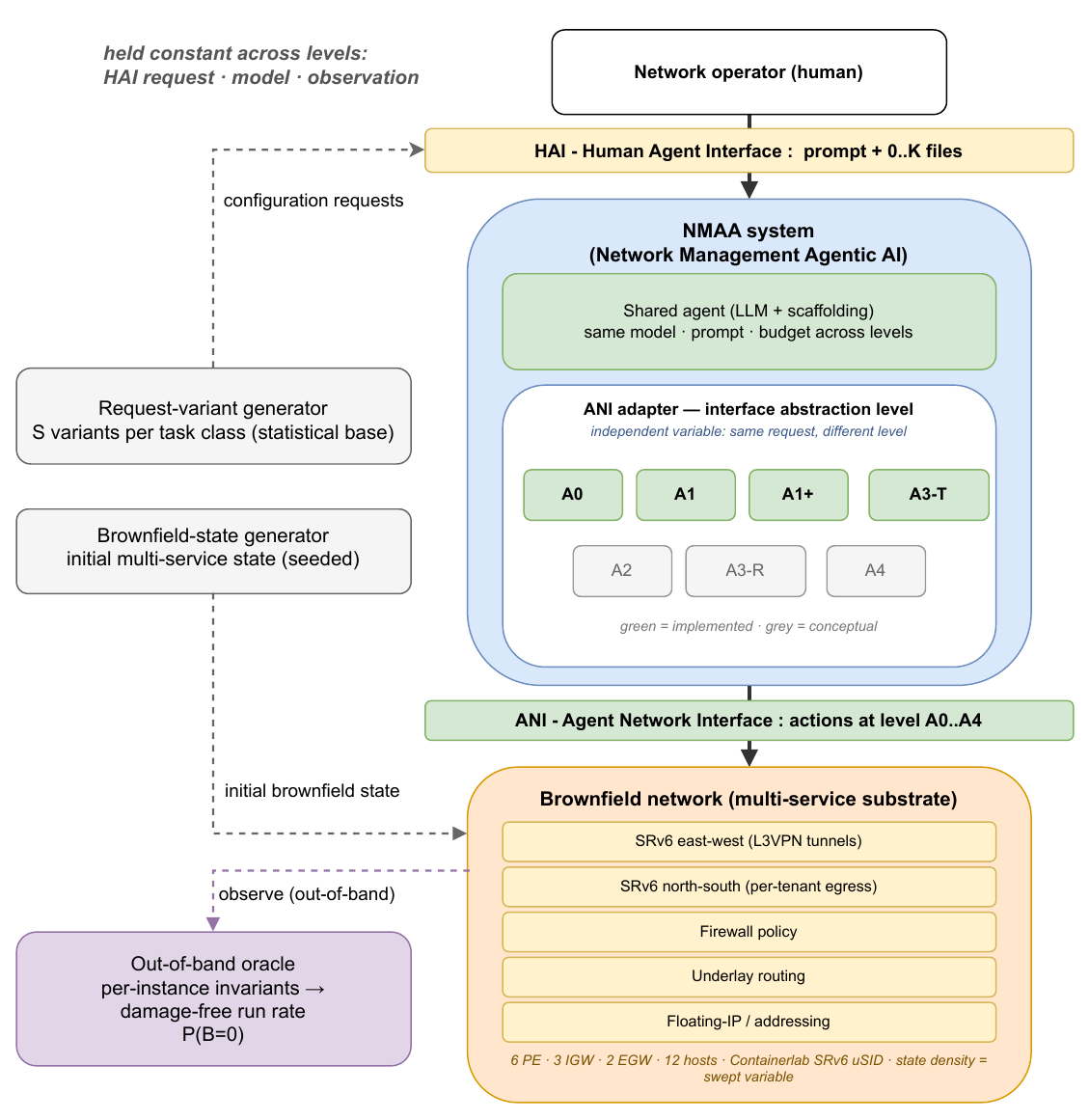}
  \caption{Detailed ANI-Gamut architecture. The shared agent acts through one ANI adapter per level (green: implemented A0/A1/A1+/A3-T; grey: conceptual A2/A3-R/A4). The request-variant generator produces the configuration requests carried over the HAI; the brownfield-state generator seeds the multi-service substrate; the out-of-band oracle checks per-instance invariants and yields the damage-free run rate $P(\mathbf{B}=\mathbf{0})$. The HAI request, the model and the observation are held constant across levels.}
  \label{fig:arch-detail}
\end{figure*}

% =====================================================================
%  Extended-only appendix: testbed RAM footprint + parallel scalability.
%  Numbers from the engineering measurement campaign (2026-06-28) and the
%  RAM footprint analysis (ani-gamut-ram-footprint-analysis.md).
%  \input from the \ifextended appendices block of main-cnsm26.tex,
%  so it appears only in the extended (arXiv) build.
% =====================================================================
\section{Testbed Footprint and Parallel Scalability}
\label{app:scalability}

A measurement campaign runs the same task across many cells (interface level, seed, state density, fault rate), so throughput is set by how many isolated worker labs run in parallel on one host.
We engineered the testbed to raise that number and, by measuring it, found that the lab footprint is not the binding constraint.
This appendix reports the progression of the per-lab RAM footprint and what actually bounds parallelism.

\paragraph{What runs per node.}
Routing is static; FRR and haproxy are present in the node image but never started, so they cost disk, not memory.
The per-router helper daemon connects to a broker that the parallel worker labs omit, so it exits a few seconds into boot and is absent during the long agent phase.
A node at steady state is therefore close to an idle shell, and the dominant per-lab cost is the Docker/containerd per-container overhead times the container count, plus one host-side agent process per worker.

\paragraph{From one container per host to an aggregated lab.}
The original layout uses one container per host: a worker lab is 23 containers (12 hosts and 11 routers).
Collapsing the twelve host endpoints into a single aggregator container, with each host leg kept in its own in-container network namespace so that the data-plane semantics are unchanged (including the isolation of overlapping-tenant addresses), brings the lab to 12 containers; the 11 routers stay separate, since each is a distinct SRv6 node and is the agent's action surface.
On the deploy VM (8 vCPU, 16\,GB) the resident cost of a lab, measured as the free-memory delta around a deployment, drops from 129\,MB to 78\,MB, about $40\%$, and the all-green baseline is reproduced identically (38 invariants, including the hard overlap case).
At roughly 5\,MB per container, however, the lab substrate sits far below the 16\,GB ceiling.

\paragraph{Estimated move to Alpine.}
The node image is currently Ubuntu with a large Python virtual environment (e.g.\ datapizza, numpy, scipy), none of which a node needs at steady state; the built image is on the order of 2--3\,GB.
A lean Alpine variant carrying only the data-plane tools (bash, iproute2, iputils-ping, nftables, iptables) would be roughly 80--150\,MB, a 15--30$\times$ reduction on disk, with faster build, pull and boot.
Its effect on resident RAM is small: only the PID-1 shell shrinks (a few MB per node), while the per-container runtime overhead that dominates is distro-independent.
Feasibility is confirmed: Alpine~3.21 ships iproute2~6.11, well above the version needed for the NEXT-CSID uSID data plane on the shared 6.8 kernel.
The Alpine image is thus a clean win on disk, build and boot, not a memory multiplier.

\begin{table}[t]
\caption{Per-lab footprint on the deploy VM (8 vCPU, 16\,GB). RAM/lab is the free-memory delta around one deployment; the Alpine row is an estimate.}
\label{tab:footprint}
\centering
\footnotesize
\begin{tabular}{@{}lccc@{}}
\toprule
\textbf{Node image / layout} & \textbf{Cont./lab} & \textbf{RAM/lab} & \textbf{Image (disk)} \\
\midrule
One container per host        & 23 & 129\,MB             & $\sim$2--3\,GB \\
Host aggregator (current)     & 12 & 78\,MB              & $\sim$2--3\,GB \\
Lean Alpine aggregator (est.) & 12 & $\sim$75\,MB        & $\sim$0.1\,GB \\
\bottomrule
\end{tabular}
\end{table}

\paragraph{What actually bounds parallelism.}
Putting the measured pieces together, a worker costs about 0.3--0.5\,GB at steady state (the lab plus one host-side agent process, hundreds of MB of Python), so memory alone would allow on the order of 25--45 parallel workers; the bursty CPU load of simultaneous redeployments caps the practical number lower, around 4--6 in transient.
Neither is the true ceiling: in practice the binding resource is the LLM API account (reachability, credit, and rate limit).
Halving the container count and shrinking the image is a clean win on deployment time and host overhead, but the lever for higher useful throughput is the API quota, not the testbed footprint.

\section{Real-World Configuration-State Scale}
\label{app:realscale}
This appendix substantiates the scale premise of Section~\ref{sec:related}: the relevant measure of playground complexity is the size and heterogeneity of the global configuration state rather than the device count.
As orders of magnitude, a production backbone or data-center network coexists on the order of $10^2$ heterogeneous service classes and carries on the order of $10^5$ to $10^6$ configuration and state items.
Table~\ref{tab:realscale} gathers public evidence, keeping \emph{service classes} (heterogeneity) separate from \emph{items per class} (density).

\begin{table}[t]
\caption{Public evidence for real-world configuration-state scale.}
\label{tab:realscale}
\centering
\footnotesize
\begin{tabular}{@{}p{3.5cm}p{2.0cm}c@{}}
\toprule
\textbf{Quantity} & \textbf{Order of magnitude} & \textbf{Source} \\
\midrule
Global BGP table (IPv4 FIB) & $\sim\!10^6$ routes ($\approx$\,950k) & \cite{huston2026bgp} \\
VRFs per device             & up to $\sim\!10^3$               & \cite{ciscoVSG} \\
BGP peers per device        & up to $\sim\!5\!\times\!10^2$    & \cite{ciscoVSG} \\
ACL entries per device      & $10^3$--$10^4$                   & \cite{ciscoVSG} \\
Configuration size per file & $10^2$--$10^3$ lines             & \cite{benson2009complexity} \\
\bottomrule
\end{tabular}
\end{table}

The density figure is well supported. A single backbone router holds on the order of $10^6$ forwarding entries, namely the global routing table~\cite{huston2026bgp}, together with up to roughly $10^3$ VRFs and $10^3$ to $10^4$ filter entries~\cite{ciscoVSG}; a network of tens to hundreds of such devices therefore reaches $10^5$ to $10^6$ distinct configuration and state items.
The heterogeneity figure is more sensitive to definition.
Counting distinct configuration feature-domains or coexisting service types, such as L3VPN, L2VPN and EVPN, Internet and peering, QoS, multicast, traffic engineering, segment routing, packet filtering, address translation, and management, and using the number of data models a platform exposes as a proxy,\footnote{Vendor YANG model inventories run to several hundred modules per release; see \url{https://github.com/YangModels/yang}.} the count lands in the tens to low hundreds, consistent with $10^2$ at the lower bound.
Counting service \emph{instances} such as per-tenant VPNs instead raises the figure to $10^2$ to $10^4$.
We thus phrase the heterogeneity as ``tens to about $10^2$ coexisting service classes'' and treat $10^2$ as an upper estimate rather than a typical value.

\section{Configuration-State Scale of Existing Agentic Benchmarks}
\label{app:arenas}
We re-examine existing agentic benchmarks on the same metric: the number of heterogeneous service classes in the initial state, and the configuration-state size per class.
Table~\ref{tab:arenas} reports figures or estimates with sources, and flags values the original work does not state.
Two groups emerge.
Network-configuration arenas run on a live or emulated substrate but expose $O(1)$ to $O(10)$ service classes with small per-class state.
Site-reliability and AIOps arenas score fault diagnosis and remediation over microservice or Kubernetes stacks; on the configuration-state metric they are off-axis, since their reported scale is a count of fault cases rather than coexisting configuration heterogeneity.

\begin{table*}[t]
\caption{Configuration-state scale of existing agentic benchmarks, measured as heterogeneous service classes against state per class. ``est.'' marks our estimate where the source is silent.}
\label{tab:arenas}
\centering
\footnotesize
\begin{tabular}{@{}lp{2.4cm}p{5.2cm}p{1.7cm}c@{}}
\toprule
\textbf{Benchmark} & \textbf{Service classes} & \textbf{State size per class} & \textbf{Substrate} & \textbf{Ref.} \\
\midrule
NIKA              & $\sim$5 scenario types  & 11--101 nodes; 54 issue types; $10^2$--$10^3$ config items (est.) & live emul.   & \cite{wang2025nika} \\
NetOpsBench       & $O(1)$ scenarios        & tens of devices, 4 scales (est.)                                & live emul.   & \cite{netopsbench2026} \\
NetArena/NetPress & 10 node types           & 5{,}493-node graph (static); live tasks small                   & static/live  & \cite{netarena} \\
NetConfEval       & 4 task families         & small topology; static translation                              & static       & \cite{wang2024netconfeval} \\
NetConfBench      & 40 tasks                & few nodes per task; intent, initial config, tests               & live emul.   & \cite{cui2025netconfbench} \\
MeshAgent         & 3 graph applications    & DSL queries over a network graph                                & graph        & \cite{meshagent2026} \\
AIOpsLab          & 2--3 microservice apps  & up to 28 microservices; 1000 fault scenarios                    & live (K8s)   & \cite{aiopslab2025} \\
Cloud-OpsBench    & K8s stack               & 452 fault cases, 40 root-cause types                            & live (K8s)   & \cite{cloudopsbench2026} \\
OperAID           & 1 app (Open5GS 5GC)     & 11 NFs + RAN sim; 3 fault scenarios; 7 read-only kubectl tools  & live (KinD)  & \cite{operaid} \\
MicroRemed        & microservice apps       & microservice topologies (est.)                                  & live         & \cite{microremed2025} \\
ITBench           & 3 IT domains            & $\sim$94 scenarios (SRE/CISO/FinOps)                            & live         & \cite{itbench2025} \\
SREGym            & cloud-native stack      & 90 SRE problems; multi-layer faults                            & live         & \cite{sregym2026} \\
\bottomrule
\end{tabular}
\end{table*}

Across both groups, no benchmark combines many heterogeneous coexisting service classes with large per-class configuration state on a live substrate.
The only entry reaching thousands of elements, NetPress, does so as a static capacity-planning graph of 5{,}493 nodes rather than a live operated network~\cite{netarena}; we therefore bind the toy-scale observation to live, interactive substrates.
For reference, the substrate of Section~\ref{sec:playground} carries on the order of $10$ coexisting service classes and $10^3$ to $10^4$ configuration items, already above the live benchmarks surveyed here, while production networks (Appendix~\ref{app:realscale}) sit two to three orders of magnitude higher again.
\fi

\section*{Acknowledgment}
    This work has received funding from the Italian MUR PRIN NEWTON project.

\bibliographystyle{IEEEtran}
\bibliography{references}

\end{document}